\documentclass[modern, twocolumn]{aastex631}

\usepackage{booktabs}
\usepackage{multirow}
\usepackage{physics}
\usepackage{tikz}
\usetikzlibrary{shapes, arrows}
\usetikzlibrary{automata,positioning,arrows,positioning}
\tikzstyle{arrow} = [thin,->,>=stealth]
\usepackage{ulem}

\let\tablenum\relax
\usepackage{siunitx}
\DeclareSIUnit\GeV{\giga\electronvolt}
\DeclareSIUnit\TeV{\tera\electronvolt}
\DeclareSIUnit\PeV{\peta\electronvolt}
\DeclareSIUnit\parsec{pc}
\DeclareSIUnit\Gpc{\giga\parsec}
\DeclareSIUnit\Mpc{\mega\parsec}
\DeclareSIUnit\erg{erg}

\NewCommandCopy{\oldeqref}{\eqref}
\renewcommand{\eqref}[1]{Eq.~\oldeqref{#1}}

\newcommand{\IRF}[1]{\ensuremath{\mathrm{IRF}(#1)}}
\newcommand{\E}[0]{\ensuremath{E}}
\newcommand{\Emin}[0]{\ensuremath{E_\text{min}}}
\newcommand{\Emax}[0]{\ensuremath{E_\text{max}}}
\newcommand{\Edet}[0]{\ensuremath{\hat{E}}}
\newcommand{\Edetmin}{\ensuremath{\hat{E}_\mathrm{min}}}
\newcommand{\Esrcmin}{\ensuremath{E^\prime_\mathrm{min}}}
\newcommand{\Edetmax}{\ensuremath{\hat{E}_\mathrm{max}}}
\newcommand{\Esrcmax}{\ensuremath{E^\prime_\mathrm{max}}}
\newcommand{\Esrc}[0]{\ensuremath{E^\prime}}
\newcommand{\Aeff}[0]{\ensuremath{A_\mathrm{eff}}}
\newcommand{\dL}[0]{\ensuremath{D_L}}
\newcommand{\Lsrc}[0]{\ensuremath{L}}
\newcommand{\dir}[0]{\ensuremath{\omega}}
\newcommand{\dirdet}[0]{\ensuremath{\hat{\omega}}}
\newcommand{\angerr}[0]{\ensuremath{\sigma_{\hat{\omega}}}}
\newcommand{\mceq}[0]{\texttt{MCEq}}
\newcommand{\stan}[0]{\texttt{Stan}}

\newcommand{\skyllh}[0]{\texttt{skyllh}}
\newcommand{\gammadiff}[0]{\ensuremath{\gamma_\mathrm{d}}}
\newcommand{\gammaps}[0]{\ensuremath{\gamma_{\mathrm{s}}}}
\newcommand{\Phips}[0]{\ensuremath{\Phi_\mathrm{s}}}
\newcommand{\Phipsi}[1]{\ensuremath{\Phi_{#1}}}
\newcommand{\Phidiff}[0]{\ensuremath{\Phi_\mathrm{d}}}
\newcommand{\Phiatmo}[0]{\ensuremath{\Phi_\mathrm{a}}}
\newcommand{\Fps}[0]{\ensuremath{F}}  
\newcommand{\phips}[0]{\ensuremath{\phi_\mathrm{s}}}  
\newcommand{\phidiff}[0]{\ensuremath{\phi_\mathrm{d}}}

\newcommand{\Nnu}[0]{\ensuremath{n}} 
\newcommand{\Nsrc}[0]{\ensuremath{N}}

\newcommand{\Nnutot}[0]{\ensuremath{n_\nu}}
\newcommand{\Nnutotexp}[0]{\ensuremath{\bar{n}_\nu^\mathrm{tot}}}
\newcommand{\Nnups}[0]{\ensuremath{n_\nu}}
\newcommand{\Nnupsi}[1]{\ensuremath{n_{#1}}}

\newcommand{\Nnupsexp}[0]{\ensuremath{\bar{n}_\nu}}
\newcommand{\Nnupsiexp}[1]{\ensuremath{\bar{n}_{#1}}}

\newcommand{\Nps}[0]{\ensuremath{N_\mathrm{s}}}
\newcommand{\Ncomp}[0]{\ensuremath{N_\mathrm{c}}}

\newcommand{\inu}[0]{\ensuremath{i}}
\newcommand{\isrc}[0]{\ensuremath{j}}

\usetikzlibrary{bayesnet}
\tikzset{<->/.tip=To}

\submitjournal{ApJ}

\shorttitle{Bayesian neutrino point source analysis}
\shortauthors{Capel et al.}

\begin{document}

\title{A hierarchical Bayesian approach to point source analysis in
  high-energy neutrino telescopes}

\correspondingauthor{Francesca Capel} \email{capel@mpp.mpg.de}

\author[0000-0002-1153-2139]{Francesca Capel}
\affiliation{Max Planck Institute for Physics \\
  Boltzmannstra\ss e 8, 85748 Garching}

\author[0009-0004-2166-6909]{Julian Kuhlmann}
\affiliation{Technical University of Munich \\
  James-Franck-Str. 1, 85748 Garching}
\affiliation{Max Planck Institute for Physics \\
  Boltzmannstra\ss e 8, 85748 Garching}

\author[0000-0003-3932-2448]{Christian Haack}
\affiliation{Erlangen Centre for Astroparticle Physics \\
  Nikolaus-Fiebiger-Str. 2, 91058 Erlangen}

\author[0000-0001-7776-4875]{Martin Ha Minh}
\affiliation{Technical University of Munich \\
  James-Franck-Str. 1, 85748 Garching}

\author[0000-0002-9566-4904]{Hans Niederhausen}
\affiliation{Michigan State University \\
  288 Farm Lane, 48824 East Lansing MI}

\author[0000-0001-8945-6722]{Lisa Schumacher}
\affiliation{Erlangen Centre for Astroparticle Physics \\
  Nikolaus-Fiebiger-Str. 2, 91058 Erlangen}

\begin{abstract}

  We propose a novel approach to the detection of point-like sources
  of high-energy neutrinos. Motivated by evidence for emerging sources
  in existing data, we focus on the characterisation and
  interpretation of these sources rather than the rejection of the
  background-only hypothesis. The hierarchical Bayesian model is
  implemented in the Stan platform, enabling computation of the
  posterior distribution with Hamiltonian Monte Carlo. We simulate a
  population of weak neutrino sources detected by the IceCube
  experiment and use the resulting data set to demonstrate and
  validate our framework. We show that even for the challenging case
  of sources at the threshold of detection and using limited prior
  information, it is possible to correctly infer the source
  properties. Additionally, we demonstrate how modelling flexible
  connections between similar sources can be used to recover the
  contribution of sources that would not be detectable
  individually. While a direct comparison of our method to existing
  approaches is challenged by the fundamental differences in
  frequentist and Bayesian frameworks, we draw parallels where
  possible. In particular, we highlight how including more complexity
  into the source modelling can increase the sensitivity to sources
  and their populations.

\end{abstract}


\section{Introduction}
\label{sec:introduction}

Neutrino astronomy is in an exciting period, with the discovery of
astrophysical neutrinos confirmed, but the search for their sources
still ongoing \citep{Kurahashi:22fg}. Recent results from the IceCube
Collaboration present evidence for the association of high-energy
neutrinos with the blazar TXS~0506+056
\citep{IceCube:2018dnn,IceCube:2018kh}, the Seyfert galaxy NGC~1068
\citep{Abbasi:2022sw} and the Galactic plane
\citep{Abbasi:2023pd}. Independent analyses making use of public
information also claim significant associations of neutrino events
with blazars \citep{Giommi:2021kd,Buson:2022ps,Buson:2023pf}, tidal
disruption events (TDEs,
\citealt{Stein:2021bl,Velzen:2021hf,Reusch:2022jk}), the Seyfert
galaxies NGC~4151 and NGC~3079 \citep{Neronov:2024dk}, and the Cygnus
region \citep{neronov2023cygnus}. However, few reports have crossed
the $5\sigma$ significance threshold typically used to define
detections and the physical interpretation of these results remains
challenging.

The approach to searching for point sources in neutrino data used by
the IceCube Collaboration makes use of hypothesis testing techniques
in a likelihood-based frequentist framework, as described in
\citet{Braun:2008bg, Braun:2010nv}. The reconstructed event
directions, energies, and angular uncertainties are used to
distinguish source and background through a likelihood ratio test,
comparing null and signal hypotheses. The significance of a potential
source location is then calculated as a $p$-value by comparing the
observed test statistic value in experimental data to the test
statistic distribution expected under the null model.

We present an alternative approach to point source searches within the
framework of Bayesian hierarchical modelling. The motivation is to
make the most of existing data. Our framework can handle fits of
complex models with large numbers of free parameters. As such, more
information from both theory and experiment can be brought together
resulting in more interpretable statistical analyses. With several
large-scale neutrino observatories either in operation
(IceCube:~\citealt{Aartsen:2017jk},
Baikal-NT:~\citealt{Belolaptikov:1997aj}), in development
(KM3Net:~\citealt{AdrianMartinez:2016bf,margiotta2022km3net};
Baikal-GVD:~\citealt{Avrorin:2021pd,Dvornicky:2023},
P-ONE:~\citealt{Agostini:2020hd}) or planned in the future
(IceCube-Gen2:~\citealt{Aartsen:2021fy},
TRIDENT:~\citealt{ye2024multicubickilometre}), we also focus on
developing methods that can adapt and scale as we learn more about
neutrino sources. As such, we focus on the characterisation of sources
in addition to their discovery.

In this work, we introduce our method in Section~\ref{sec:methods} and
demonstrate its performance on simulated data in
Section~\ref{sec:application}. We then discuss the performance of our
framework in the context of existing methods in
Section~\ref{sec:discussion} before concluding in
Section~\ref{sec:conclusions}. The code used in this work and relevant
examples are available in the \texttt{hierarchical\_nu} python package
\citep{Capel:2024gh}.

\section{Methods}
\label{sec:methods}

Our approach is centred on the derivation of a hierarchical or
multi-level likelihood function that captures the key phenomenology of
astrophysical neutrino production and detection. We start with the
high-level model parameters and connect to the observables: the
neutrino energies and arrival directions, along with their respective
reconstruction uncertainties. We describe the key aspects of the
physical model and the statistical implementation here.

\subsection{Physical model}
\label{sec:physical_model}

We consider point-like sources of astrophysical neutrinos with a
power-law spectrum defined between source frame energies \Esrcmin{}
and \Esrcmax{}
\begin{equation}
  \frac{\dd[2]{\Nnu{}}}{\dd{E}\dd{t}} \propto E^{-\gammaps{}},
\end{equation}
where \gammaps{} is the spectral index. The isotropic source
luminosity with this energy range leads to an energy flux at Earth
\begin{equation} \label{eq:flux_thinning} \Fps{} =
  \frac{L}{4\pi\dL{}^{2}(z)}
\end{equation}
in the redshifted energy range
$E_{\text{min, max}} = \Esrc{}_{\text{min, max}} / (1+z)$, where
$\dL{}(z)$ is the luminosity distance at redshift $z$.  It then
follows that the differential flux from a single source at Earth is
given by
\begin{equation}\label{eq:ps_flux}
  \frac{\dd[3]{\Nnu{}}}{\dd{E}\dd{t}\dd{A}} = \frac{L k_\gamma E^{-\gammaps{}}}{4 \pi D_L(z)^2} = \phi_\mathrm{s} \Bigg(\frac{E}{E_0}\Bigg)^{-\gammaps{}},
\end{equation}
where $k_\gamma$ is defined such that the spectrum is normalised to
$L$ in the source frame and in the final step we introduce the
normalisation energy, $E_0$, and summarise the differential flux at
$E_0$ as \phips{}. Throughout this work, we assume a flat $\Lambda$CDM
cosmology with $H_0 = 70$~\si{\kilo\metre\per\second\per\Mpc},
$\Omega_m = 0.3$ and $\Omega_\Lambda = 0.7$. We choose a standard
power-law spectral model for point sources here to allow for
straightforward comparison to other methods. However, more complex
modelling possibilities are implemented in our framework and we plan
to explore their application in future work.

In addition to point sources, we also consider two diffuse components
as possible sources of neutrino emission: a diffuse astrophysical
background and an atmospheric background.

Individual sources of interest typically belong to a population of
objects with similar astrophysical properties. In point-source
searches, we typically only hope to resolve some fraction of the total
population of sources. Additionally, individual sources or lists of
sources may not be solely responsible for the observed astrophysical
flux. With these factors in mind, we model a diffuse astrophysical
component that accounts for any astrophysical flux that cannot be
associated with resolved point sources. To avoid assumptions on the
cosmological distribution or luminosities of these unknown sources, we
simply model this contribution as an isotropic spectrum over the whole
sky described as
\begin{equation}\label{eq:isotropic_flux}
  \frac{\dd[4]{\Nnu{}}}{\dd{E}\dd{t}\dd{A}\dd{\dir{}}} = \phidiff{} \Bigg( \frac{E}{E_0} \Bigg)^{-\gammadiff{}},
\end{equation}
where \phidiff{} is the differential flux normalisation at $E_0$, and
\gammadiff{} is the spectral index of the bounded power law spectrum
that is defined between \Emin{} and \Emax{}. This spectral model is
standard in diffuse astrophysical neutrino analyses, but could be
adapted as new information becomes available (see
e.g.~\citealt{Naab:2023V2}).

Another important diffuse background, particularly at energies
$E < 100$~\si{\TeV}, is that due to atmospheric neutrinos, produced by
the interactions of cosmic rays in our Earth's atmosphere. The
atmospheric neutrino flux depends on the zenith angle and the spectrum
is not well-described by a single power law. Therefore, we use
\mceq{}~\citep{fedynitch2015calculation} to model the atmospheric
arrival direction distribution and spectra. We use the H4a cosmic ray
flux model described in \citet{Gaisser:2012hf}, the atmospheric
density profile implemented in \texttt{NRLMSISE-00}
\citep{Picone:2002yd} and \texttt{SIBYLL~2.3c} to model hadronic
interactions \citep{Riehn:2017ud}. Furthermore, we assume that the
normalisation of the atmospheric flux is not exactly known and
parameterise it as \Phiatmo{}.

High-energy neutrino telescopes measure the secondary Cherenkov
radiation produced when incoming neutrinos interact in a large
instrumented volume of water or ice. The energies and arrival
directions of the primary neutrinos are not observable, but the
reconstructed energies and directions of secondaries serve as a
proxy. The reconstruction of these observables yields an associated
uncertainty. Here, we consider the IceCube neutrino observatory to
demonstrate our approach, but it is straightforward to extend our
framework to other experiments. In particular, we consider the
publicly available data of track-like events from muon neutrino
interactions \citep{icecube_data} that is used in
\citet{icecube_point_source}, but with limited information on the
provided instrument response functions (IRFs). The IRFs consist of the
effective area, \Aeff{}, as function of neutrino energy, \E{}, and
declination, $\delta$. Further, the energy resolution and point spread
function are available as a tabulated mapping from \E{} and $\delta$
to \Edet{} and \dirdet{}. The corresponding public data set lists
reconstructed muon events with energy \Edet{}, and direction
\dirdet{}. Further details are discussed in
Appendix~\ref{sec:neutrino_detection}.

\subsection{Statistical formalism}
\label{sec:statistical_formalism}

The aim of our statistical framework is to quantify the association of
neutrinos with possible sources, whilst simultaneously inferring the
physical properties of these sources. We expect sources to share
similar characteristics across their classes or populations, making a
hierarchical framework a natural approach.

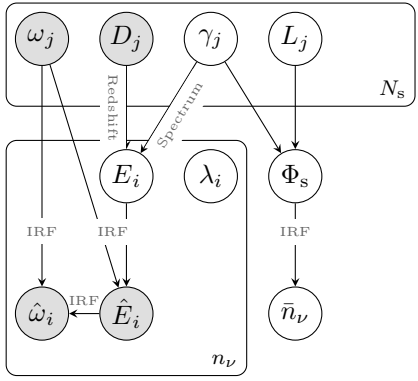
\begin{figure}[h]
  \centering
  \begin{tikzpicture}
    \matrix[row sep=0.55cm, column sep=0.4cm] (LDA) { %
      \node[obs] (omega) {$\dir{}_\isrc{}$}; & %
      \node[obs] (D) {$D_\isrc{}$}; & %
      \node[latent] (gamma) {$\gamma_\isrc{}$}; & %
      \node[latent] (L) {$L_\isrc{}$}; & %
      \node[latent, draw=none] (empty) { }; & %
      \\
      \\
      & %
      \node[latent] (E_arr) {$\E{}_\inu{}$}; & %
      \node[latent] (lambda) {$\lambda_\inu{}$}; & %
      \node[latent] (Phi_s) {$\Phips{}$}; & %
      & %
      \\
      \\
      \node[obs] (omega_det) {$\dirdet{}_\inu{}$}; & %
      \node[obs] (E_det) {$\Edet{}_\inu{}$}; & %
      & %
      \node[latent] (Nnupsexp) {$\Nnupsexp{}$}; & %
      & %
      \\
    }; \plate {sources} {(gamma) (D) (omega) (L) (empty)}
    {$\Nps{}$}; %
    \plate {nu} {(E_arr) (E_det) (lambda) (omega_det)} {$\Nnutot$}; %
    \draw [arrow] (gamma) --
    node[midway,sloped,below,color=black!60,fill=white]
    {\tiny{Spectrum}} (E_arr); %
    \draw [arrow] (D) --
    node[midway,sloped,below,color=black!60,fill=white]
    {\tiny{Redshift}} (E_arr); %
    \draw [arrow] (omega) -- node[midway,sloped,above,color=black!60]
    {} (E_det); %
    \draw [arrow] (E_arr) --
    node[midway,above,color=black!60,fill=white,xshift=-0.18cm]
    {\tiny{IRF}} (E_det); %
    \draw [arrow] (omega) --
    node[above,color=black!60,fill=white,pos=0.81] {\tiny{IRF}}
    (omega_det); %
    \draw [arrow] (E_det) -- node[midway,sloped,above,color=black!60]
    {\tiny{IRF}} (omega_det); %
    \draw [arrow] (L) -- node[midway,sloped,above] {} (Phi_s); %
    \draw [arrow] (gamma) -- node[midway,sloped,above] {} (Phi_s); %
    \draw [arrow] (Phi_s) --
    node[midway,above,color=black!60,fill=white] {\tiny{IRF}}
    (Nnupsexp); %
  \end{tikzpicture}
  \caption{Summary of the likelihood for point source components. Open
    and shaded circles are used to show model parameters and fixed
    observables respectively, with the arrows showing the connections
    between them. The boxes are used to separate the \Nps{} source
    compenents and \Nnutot{} events. The parameters are explained in
    the text. The levels from upper to lower show the source
    properties, latent or ``hidden'' true parameters that depend on
    the source properties and cannot be directly observed, and
    observable quantities that depend on the IRF. The latent source
    labels, $\lambda_i$, are discrete parameters marking the mixture
    model component that each event belongs to.}
  \label{fig:graphical_model}
\end{figure}

Our method builds on a hierarchical mixture model formalism for
cross-identification (see e.g. \citealt{Budavari:2015hj}), extended to
account for reconstructed event energies as well as directions,
important selection effects in the observed samples, and relevant
source properties \citep{Capel:2019mv}. The likelihood has the form of
an inhomogeneous Poisson point process, the rate of which is a mixture
model over the different possible source contributions. This can be
written as \citep{Streit:2010}

\begin{equation}
  \mathcal{L}(\E{}, \Theta) = e^{-\Nnutotexp{}} \prod_{\inu{}=1}^{\Nnutot} \sum_{\isrc{}=1}^{\Ncomp} r_\isrc{}(\Edet{}_\inu{}, \dirdet{}_\inu{} | \E{}_\inu{}, \Theta_\isrc),
  \label{eq:likelihood}
\end{equation}
for \Nnutot{} detected neutrino events and \Ncomp{} model components
(where \Ncomp{} = \Nps{} + 2 diffuse components). The expected number
of detected neutrino events, \Nnutotexp{}, is given by the sum over
all model components such that
${\Nnutotexp{} = \sum_{\isrc{}=0}^{\Ncomp{}}
  \Nnupsiexp{\isrc}}$. Similarly, the expected number of events from
point sources is given by
${\Nnupsexp{} = \sum_{\isrc{=0}}^{\Nps{}} \Nnupsiexp{\isrc}}$. We
summarise the vector of all high-level parameters as $\Theta$, and the
subset of those relevant for individual model components as
$\Theta_\isrc{}$. All energies are summarised as $\E{}$. For \Nps{}
point sources,
${\Theta_\isrc{} = \{L_\isrc{}, \gamma_\isrc{}, \dir{}_\isrc{},
  D_\isrc{} \}}$, whereas for the diffuse astrophysical and
atmospheric components
${\Theta_\isrc{} = \{ \Phidiff{}, \gammadiff{} \}}$ and
${\Theta_\isrc{} = \{ \Phiatmo{} \}}$, respectively. $r_\isrc{}$ is
the \isrc{}-th component's rate parameter, relating to the number of
expected events, \Nnutotexp{}, through integration as shown in
\eqref{eq:app_nex}. The relationship between parameters in
\eqref{eq:likelihood} is summarised graphically in
Fig.~\ref{fig:graphical_model} for the case of point source
components.

All model parameters are left free, but the highest-level parameters
also have associated prior distributions. As a default, we make use of
weakly informative priors for these parameters, as shown in
Table~\ref{tab:priors}. Such priors allow us to include our knowlegde
on reasonable physical values, while not driving the results of the
eventual inference. In Section~\ref{sec:multi_source}, we also
demonstrate the application of more informative priors, as could be
motivated by, e.g., multi-messenger information or theoretical
predictions. Where relevant, we verify that the results are robust to
reasonable variations in the choice of priors.

\begin{table}[ht]
  \centering
  \begin{tabular}{ccc}
    \toprule
    & Distribution & Units \\
    \cmidrule{2-3}
    $L/L_\isrc{}$ & $\mathrm{Lognorm}(8.0 \times 10^{43}, 4.0)$ & \si{\erg\per\second}\\
    $\gammaps{}/\gamma_\isrc{}$ & $\mathrm{Norm}(2.0, 0.25)$ & -- \\
    $\Phidiff{}$ & $\mathrm{Lognorm}(5.4 \times 10^{-8}, 0.3)$ &  \si{\per\centi\meter\squared\per\second} \\
    $\gammadiff{}$ & $\mathrm{Norm}(2.5, 0.04)$ & -- \\
    $\Phiatmo{}$ & $\mathrm{Lognorm}(3.0 \times 10^{-5}, 0.08)$ & \si{\per\centi\meter\squared\per\second} \\   
    \bottomrule
  \end{tabular}
  \caption{Default prior assumptions for model hyperparameters, given
    as the distributions used and the corresponding $\mu$ and $\sigma$
    values. $L/L_\isrc{}$ is bounded between
    $[0, 10^{52}]$~\si{\erg\per\second}, \Phidiff{} and \Phiatmo{} are
    also bounded between $[10^{-10}, 10^{-7}]$ and
    $[0, 3 \times
    10^{-4}]$~\si{\per\centi\metre\squared\per\second}. Both
    $\gammaps{}/\gamma_\isrc{}$ and \gammadiff{} are bounded between 1
    and 4.}
  \label{tab:priors}
\end{table}

For a given data set of \Nnutot{} neutrino events, we perform
inference on the model parameters in a Bayesian framework. The
posterior distribution is proportional to the likelihood given in
\eqref{eq:likelihood} multiplied by the joint prior summarised in
Table~\ref{tab:priors}. We generate samples from this posterior using
\stan{}~\citep{Stan:2023pf}, which implements an efficient variant of
a Markov chain Monte Carlo algorithm called Hamiltonian Monte
Carlo. This class of algorithm guarantees convergence to the target
distribution in the limit of infinite samples. Here, we use a set of
diagnostics to assess convergence in the case of finite samples. In
particular, we run 4 separate chains of 2000 samples each (1000
warm-up samples to tune the algorithm and find the target
distribution, 1000 actual samples of the target distribution) and
require an effective sample size, $n_\mathrm{eff} > 1000$ and a
Gelman-Rubin statistic, $\hat{R} < 1.1$ with no divergent transitions
for all model parameters in our analyses~\citep{Gelman:2013pd}.

\subsection{Interpretation of results}
\label{sec:interpretation}

The results of the fits using \stan{} take the form of samples
representing the joint posterior distribution over all model
parameters, also including latent parameters. Useful summaries of the
fit parameters can be derived from these samples, such as the most
probable value and highest posterior density interval (HDI),
representing the ``best-fit'' value and its associated uncertainty. We
evaluate the ``goodness-of-fit'' using posterior predictive checks
(PPCs), which involve generating simulated data under the assumptions
of the fitted model and comparing to the observed data to check the
inferences are reasonable \cite[Chapter 6.3]{Gelman:2013pd}.

As shown in Fig.~\ref{fig:graphical_model}, each neutrino event has a
latent true energy parameter, $\E{}_\inu{}$. The marginal posterior of
this parameter provides additional information useful to the
interpretation of the origin of this event, considering all possible
model components and uncertainties, which is non-trivial to infer from
the reconstructed event energies, $\Edet{}_\inu{}$, alone.

To directly address the possible association of neutrino events with
source components of the model, we introduce additional derived
parameters that capture the relevant information. Each neutrino event
has a discrete label parameter, $\lambda_\inu{}$, which identifies its
source component with possible values in the range
$[1,\Ncomp{}]$. While the $\lambda_\inu{}$ are not explicity sampled
during inference, we can compute the marginal posterior distribution
for $\lambda_\inu{}$
\begin{equation}
  \Pr(\lambda_\inu{} = \isrc{} \vert \Edet{}, \dirdet{}, \Theta) =
  \frac{r_\isrc{}(\Edet{}, \dirdet{} \vert \Theta_\isrc{})}
  {\sum_{l=1}^{\Ncomp{}}r_l(\Edet{}, \dirdet{} \vert \Theta_{l})}.
\end{equation}
This probability is used as the colour scale in
Fig.~\ref{fig:single_source_roi}. More details can be found in
\citet[Chapter 3.2]{Streit:2010} and the \stan{}
documentation\footnote{\url{https://mc-stan.org/docs/stan-users-guide/finite-mixtures.html\#recovering-posterior-mixture-proportions}}. This
marginal posterior gives the association probability for each
event--source component pair given the available data\footnote{For any
  event, the sum of the association probabilities over all possible
  source components is equal to 1.}, allowing for a more direct and
insightful intepretation of the results.

We can also quantify discovery and sensitivity within this Bayesian
framework. For individual event--source associations, we can
confidently claim an association if
\begin{equation}\label{eq:discovery_event}
  \Pr(\lambda_\inu{} = \isrc{} | \Edet{}, \dirdet{}) > \alpha,
\end{equation} where $\alpha$ is a threshold probability to be defined. However, the information in individual events is naturally
limited, and therefore if using small event sample sizes it is
important to study the prior dependence of the association probability
(see \citealt{Capel:2023pf} for an example). A useful parameter that
summarises our expectation over the whole sample of possible
event--source associations is the expected number of detected
neutrinos from the \isrc{}-th point source, \Nnupsiexp{\isrc}. We
define the detection of a source as
\begin{equation}\label{eq:discovery_nex}
  \Pr(\Nnupsiexp{\isrc} \geq 1 | \Edet{}, \dirdet{}) > \alpha.
\end{equation}
In the case of a non-detection or when quantifying the sensitivity, we
can find the HDI of \Phipsi{\isrc} or $L_\isrc$ for a given
probability level, $\alpha$, and report the corresponding upper limit.

The probability thresholds, $\alpha$, defined above are somewhat
arbitrary and are not directly comparable to a $p$-value from a
frequentist analysis. It is possible to calibrate $\alpha$ via
repeated simulations to give some desired coverage, true detection
rate or false detection rate
\citep{betancourt2018calibrating}. However, within a Bayesian
framework, the resulting posterior probability for event--source
associations or expected number of events is itself the main
result. For the purpose of this work we choose $\alpha = 0.95$
(similar definitions can be found in, e.g.,
\citealt{Aggarwal:2021fk,Sottosanti:2021kl}). This is not equivalent
to a $p$-value of 0.05, and is importantly a statement about the
alternate hypothesis (point-source model components) rather than the
null hypothesis (diffuse background model components).

\section{Application}
\label{sec:application}

In this work, we focus the application of our method to a population
of weak sources that are below the detection threshold of existing
methods.

\subsection{Simulated data set}
\label{sec:sim_data}

We simulate a population of neutrino sources using \texttt{popsynth}
\citep{Burgess:2021kg}, characterised by the physical properties
introduced in Section~\ref{sec:physical_model}: a redshift, $z$, a
luminosity, $L$, a spectral index, $\gamma$, and a direction,
$\omega$. We choose a source redshift evolution based on the shape of
the star formation rate given in~\cite{Madau:2014mv} and a luminosity
function that is described by a broken power law. Source spectral
indices follow a Gaussian distribution and sources are disributed
isotropically on the sky. The choices regarding the source population
properties are detailed in Table~\ref{tab:source_pop}.

\begin{table}[ht]
  \centering
  \begin{tabular}{cc}
    \toprule
    & Distribution \\
    \cmidrule{2-2}
    $z$ & $\frac{\dd{\Nsrc}}{\dd{V}} = 100 \frac{1 + 4.8 z}{1 + (z/2.7)^{3.9}} $~\si{\per\cubic\Gpc\steradian} \\
    \multirow{3}{*}{L} & \multirow{3}{*}{$\frac{\dd{\Nsrc}}{\dd{L}} =  \begin{cases} C L^{-2} & \mbox{if } L \leq L_\mathrm{br}\\ C L^{-3} L_\mathrm{br} & \mbox{if } L > L_\mathrm{br} \end{cases}$~\si{\per\erg\second}} \\
    & \\
    & \\
    $\gamma$ & $\frac{\dd{\Nsrc}}{\dd{\gamma}} = \mathrm{Norm}(2.0, 0.25)$ \\
    $\omega$ & $\frac{\dd{\Nsrc}}{\dd{\omega}} = \frac{1}{4\pi}$~\si{\per\steradian} \\
    \bottomrule
  \end{tabular}
  \caption{Source population properties and their distributions. We
    use $L_\mathrm{br} = 5 \times 10^{44}$~\si{\erg\per\second} and
    $C$ is defined such that the distribution is normalised to 1. The
    the following ranges are also used: $z \in [0, 6]$,
    $L \in [5 \times 10^{43}, 5 \times 10^{48}]$~\si{\erg\per\second}
    and $\gamma \in [1, 4]$.}
  \label{tab:source_pop}
\end{table}

Our source population is set up to provide a generic but relevant test
scenario that is not connected to a particular class of astrophysical
sources. We ensure that our simulated population is consistent with
general population constraints \citep{Murase:2016gly,Capel:2020qw} in
that it does not exceed the measured diffuse astrophysical neutrino
flux and sources are neither too rare nor too bright. Beyond
$z \sim 0.5$, the neutrino signal from the population is effectively
diffuse ($\Nnupsexp \lesssim 1$ for individual sources).

To generate our simulated neutrino data set, we select sources in the
Northern hemisphere ($\delta > -5^\circ$) with
$\Fps \geq 5 \times
10^{12}$~\si{\erg\per\square\centi\metre\per\second}, giving
$\Nps = 6$. The properties of these selected sources are given in
Table~\ref{tab:multi_src_truth}.

\begin{table*}[ht]
  \centering
  \begin{tabular}{ccccccccc}
    \toprule
    & $(RA, \delta)_\isrc$ & $z$ & $L_\isrc$ & $\gamma_\isrc$ & \multicolumn{2}{c}{\Nnupsi{\isrc{}}} & \multicolumn{2}{c}{\Nnutot{} in $5^\circ$ ROI} \\
    \cmidrule{6-9}
    & [\si{\degree}] & & [\si{\erg\per\second}] & & $\geq 300$~\si{\GeV} & $\geq 1$~\si{\TeV} & $\geq 300$~\si{\GeV} & $\geq 1$~\si{\TeV} \\
    \cmidrule{1-9}
    1 & (183.6, 5.5) & 0.04 & $5.8 \times 10^{43}$ & 2.05 &  14 & 11 & 3535 & 1989 \\
    2 & (56.0, 29.3) & 0.43 & $5.8 \times 10^{45}$ & 2.30 & 7 & 6 & 2203 & 1141 \\
    3 & (273.2, 28.5) & 0.13 & $5.3 \times 10^{44}$ & 2.52 & 5 & 5 & 2711 & 1422 \\
    4 & (23.8, 15.4) & 0.06 & $6.6 \times 10^{43}$ & 1.99 & 4 & 4 & 2245 & 1134 \\
    5 & (179.6, 20.8) & 0.17 & $1.6 \times 10^{45}$ & 1.62 & 2 & 2 & 2522 & 1273 \\
    6 & (325.0, 62.3) & 0.12 & $1.81 \times 10^{44}$ & 1.81 & 1 & 1 & 1594 & 789 \\
    \bottomrule
  \end{tabular}
  \caption{Properties of the simulated neutrino point sources. Sources
    are listed in order of decreasing number of simulated events,
    \Nnupsi{\isrc{}}, that are included in the event selection used
    here. The total number of simulated neutrino events within a
    $5^\circ$ ROI around each source is also shown for reference.}
  \label{tab:multi_src_truth}
\end{table*}

\begin{figure}[ht]
  \centering
  \includegraphics[width=\columnwidth]{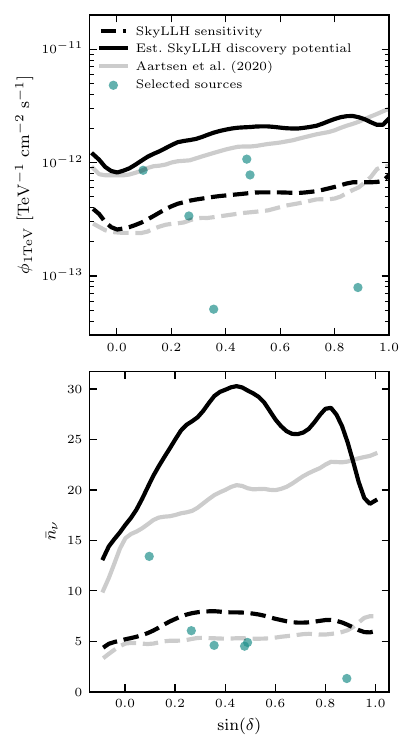}
  \caption{\textbf{Upper panel:} The differentital flux normalisation
    at 1~\si{\TeV} for simulated point sources. The \texttt{SkyLLH}
    sensitivity assuming an $E^{-2}$ power-law spectrum as reported in
    \citet{Bellenghi:20230u} is shown for comparison, as this
    calculation makes use of the same publicly available IRF that is
    used here. We also plot an estimation of the discovery potential
    by rescaling the sensitivity as a function of declination. This
    rescaling is calculated as the ratio of the discovery potential to
    the sensitivity, according to the results presented in
    \citet{Aartsen:2020ld}, which are also shown for
    comparison. \textbf{Lower panel:} The total expected number of
    neutrino events for simulated point sources is shown in comparison
    to that required by the sensitivity and discovery potential curves
    introduced above.}\label{fig:src_sensitivity}
\end{figure}

Fig.~\ref{fig:src_sensitivity} shows the selected sources in
comparison with the sensitivity and discovery potential of \skyllh{},
which implements a frequentist approach that is typically used in
neutrino point source searches
\citep{Wolf:2019Pu,Bellenghi:20230u}. The sensitivity flux is
calculated as the median 90\% CL upper limit on the flux normalisation
over repeated simulations of only background. The discovery potential
is defined as the flux that leads to a $5\sigma$ deviation from the
background expectation in 50\% of repeated simulations. The same
curves from \citet{Aartsen:2020ld} are also shown, but as these have
been calculated using more detailed detector Monte Carlo that is not
publicly available, they do not represent a fair comparison in this
work.

All sources are below the estimated discovery potential, but it is
important to note that the sensitivity and discovery potential fluxes
are calculated for a fixed power-law spectrum with index
$\gammaps{}=2$, while the point sources simulated here have
$\gamma_\isrc$ in the range $[1.62, 2.52]$. In
Fig.~\ref{fig:src_sensitivity}, we also show the corresponding number
of events in our simulation. This visualiation better accounts for the
different spectral index assumptions, but the number of events
required for a detection in a given simulation will depend on their
energies. In this work, we only investigate one possible realisation
of our simulation and Fig.~\ref{fig:src_sensitivity} is only intended
to provide a rough comparison with existing work.

\begin{figure*}[!ht]
  \centering
  \includegraphics[width=0.64\textwidth]{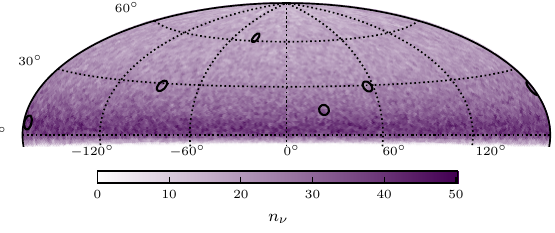}
  \includegraphics[width=0.34\textwidth]{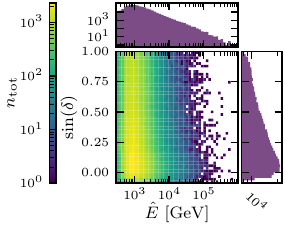}
  \caption{The simulated data set containing 669,007 events with
    $\hat{E} > 300$~\si{GeV}. The left panel shows a skymap of the
    event directions with the selected source positions overlaid. The
    right panel shows the distribution of events in $\Edet$ and
    $\sin(\delta)$, cf. Fig.~1 in
    \citet{icecube_data}.}\label{fig:sim_summary}
\end{figure*}

All other sources from the population are treated as a sub-dominant
contribution to the diffuse flux and are not simulated explicitly. We
include a diffuse astrophysical flux with
$\phidiff = 1.8 \times
10^{-18}$~\si{\per\GeV\per\square\centi\metre\per\second\per\steradian}
and $\gammadiff = 2.5$ based on recently reported observations
\citep{Abbasi:2022kf,Naab:2023V2,Abbasi:2024fd}. The atmospheric flux
is also included using \mceq\ as described in
Section~\ref{sec:physical_model}. All source components are simulated
over a wide energy range from $\Emin = 100$~\si{\GeV} to
$\Emax = 100$~\si{\PeV}. We simulate the detection of neutrinos from
these source components using the provided IRF and place a cut on the
reconstructed muon energy of $\Edetmin = 300$~\si{\GeV}. This choice
is to include thresholding effects due to events with true energies
$\E{} < 300$~\si{GeV} being mis-reconstructed above \Edetmin{} and
vice versa. For this simulated data set, we simulate \Nnupsi{\isrc{}}
as the most likley integer value based on \Nnupsiexp{\isrc{}}, for a
clearer interpretation and discussion of the different source cases in
the next sections. A summary of the resulting data set is shown in
Fig.~\ref{fig:sim_summary}.

Here, we only analyse a single realisation of the simulated data set
for clarity. However, in developing our method we verified that it
performs well and without biases by analysing 100s of realisations of
test point sources at a range of declinations. We also tested the
performance of our definition of source detection detailed in
Section~\ref{sec:interpretation}, finding no false detections in
background-only data sets.

\subsection{Single source analysis}
\label{sec:single_source}

To demonstrate the analysis of individual sources in our framework we
focus on the point source that has the most detected events in our
simulated data set, source~\#1 from
Table~\ref{tab:multi_src_truth}. We make a selection on our data set,
defining a circular region of interest (ROI) on the sky centred on the
source location with a radius of $5^\circ$\footnote{We verified that
  for the source modelling assumptions used here and the angular
  resolution provided by the IRF, this radius is large enough to avoid
  impacting the results.}.

\begin{figure}
  \centering \includegraphics[width=\columnwidth]{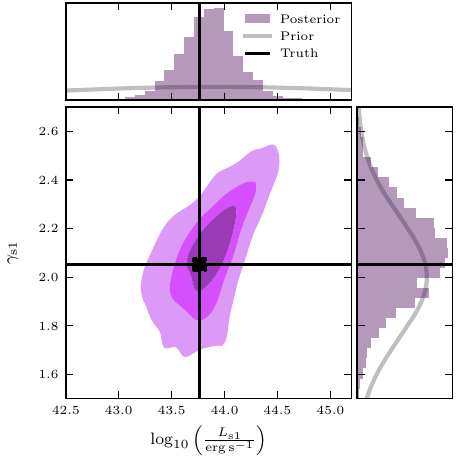}
  \caption{The joint posterior density of \Lsrc{} and \gammaps{} with
    contours showing the 30\%, 60\% and 90\% HDIs. The true parameter
    values are indicated by the horizontal and vertical solid
    lines. The upper and right panels show the marginal distributions
    in comparison with the prior.}
  \label{fig:single_source_L_src_index}
\end{figure}

We use our method to fit the data in this ROI, using the priors
detailed in Table~\ref{tab:priors}. We show the results for the
marginal posterior of the source parameters in
Fig.~\ref{fig:single_source_L_src_index}, demonstrating the correct
reconstruction of the known true parameter values.

\begin{figure}
  \includegraphics[width=\columnwidth]{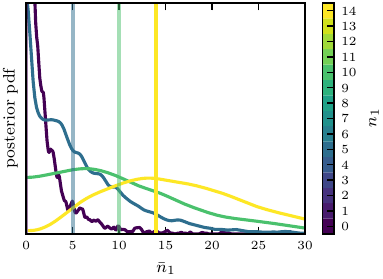}
  \caption{Marginal posterior density of the expected number of point
    source events for source \#1, \Nnupsiexp{1}. The number of
    selected source events in each fit is colour coded, with
    \Nnupsi{1} = [0, 5, 10, 14] shown.}
  \label{fig:single_source_varying_N}
\end{figure}

The marginal posterior of the expected number of neutrino events from
source \#1, \Nnupsiexp{1}, is given in
Fig.~\ref{fig:single_source_varying_N}. To validate the correct
reconstruction of \Nnupsiexp{1}, we also fit data sets where we add in
source events one by one. Fig.~\ref{fig:single_source_varying_N} also
shows the resulting \Nnupsiexp{1} posterior for
$\Nnups{} = [0, 5, 10, 14]$. We can see that in the case of
$\Nnups{} = 0$, the \Nnupsiexp{1} posterior is strongly peaked at zero
and gradually moves away in a consistent manner as more point source
events are added to the simulation. Using the definition of source
detection given in \eqref{eq:discovery_nex} with $\alpha=0.95$, we can
identify this source above the detection threshold in the simulated
dataset that includes all 14 source events. We also performed fits to
100 simulations of background-only events at this source declination,
verifying that none of these result in false point-source detections
according to our detection criterion.

\begin{figure*}
  \centering
  \includegraphics[width=\textwidth]{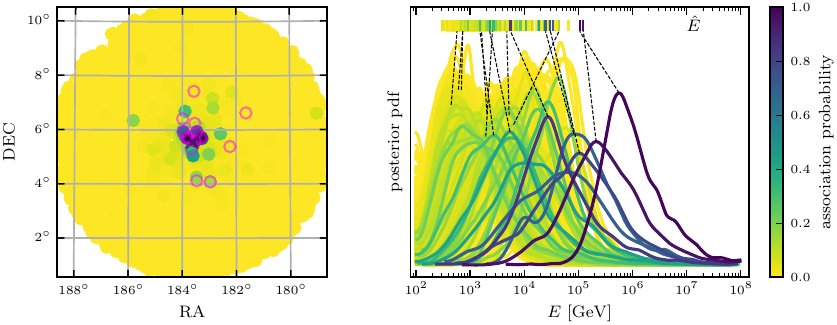}
  \caption{Colour-coded posterior association probabilities of events
    to source \#1. \textbf{Left panel:} Spatial distribution of
    events. The marker size does not represent the angular resolution
    and the point source events are highlighted in red
    borders. \textbf{Right panel:} Marginal posteriors for the true
    neutrino energies of events included in the ROI. Markers on the
    top row indicate \Edet{} for all events included in the analysis,
    with dashed connecting lines to the posteriors for the events from
    the tested point source.}
  \label{fig:single_source_roi}
\end{figure*}

Our framework further enables us to give posterior association
probabilities of single events to each source component, as detailed
in Section~\ref{sec:interpretation}. We summarise this information in
Fig.~\ref{fig:single_source_roi} in terms of both the event positions
and energies. We see that the higher energy events that are closer to
the source have a larger association probability, as expected. Point
source events that have a low association probability are mostly
mis-reconstructed as belonging to the atmospheric background
component, which dominates at lower energies. Using the event-based
definition of association given in \eqref{eq:discovery_event}, only
the event with the second highest \Edet{} value passes the threshold
with an association proability of 0.99. The event with the highest
\Edet{} value is close behind, with an association probability of
0.95.

In Fig.~\ref{fig:single_source_roi}, we see the complementary
information that is available in terms of the individual event--source
association probabilities. For example,
Fig.~\ref{fig:single_source_varying_N} shows that the overall
\Nnupsiexp{1} posterior is consistent with 14 events for source \#1,
but Fig.~\ref{fig:single_source_roi} shows that we have probable
individual associations for only about half of these events, typically
at higher energies. The remaining contribution to \Nnupsiexp{1} comes
from many events with smaller association probabilities. We can also
see that for events that have a $\sim 50\%$ association probability,
the posterior for \E{} can have two peaks, corresponding to the
possible latent \E{} that would best fit the source or background
components.

\begin{figure*}
  \includegraphics[width=\textwidth]{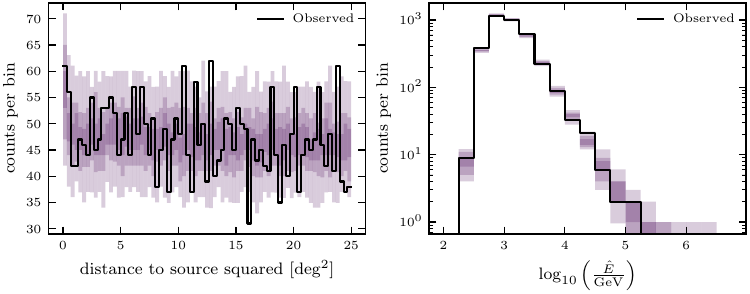}
  \caption{Distribution of observed and posterior predictive data.
    Shaded bands show 30\%, 60\%, and 90\% quantiles from darkest to
    lightest.}
  \label{fig:single_source_ppc}
\end{figure*}

As means of verifying goodness-of-fit we perform PPCs. Histograms of
\Edet{} and \dirdet{} for 100 generated data sets from the posterior
predictive distribution are shown in Fig.~\ref{fig:single_source_ppc},
indicating no obvious signs of mismodelling from visual inspection. We
also quantified the discrepancy in each bin by calculating a
$p$-value, and found only relatively large $p$-values that followed a
uniform distribution between 0 and 1.

\subsection{Joint analysis of multiple sources}
\label{sec:multi_source}

We now perform joint fits of all six selected point sources. Motivated
by computational considerations, we make a further cut of
$\Edet{} \geq 1$~\si{\TeV} on our simulated dataset, as only 4 of the
total 33 point source events are lost in this case (see
Table~\ref{tab:multi_src_truth}). Additionally, based on the results
in Section~\ref{sec:single_source}, we do not expect the lower energy
events to contribute to the detectability of these sources. Our
framework is designed to allow for flexible definitions of possible
connections between individual sources, and we explore three cases
here as examples: 1) All sources share the same luminosity, $L$, and
spectral index, \gammaps{}; 2) All sources have individual and unknown
$L_\isrc{}$ and $\gamma_\isrc{}$; and 3) All sources have individual
$L_\isrc{}$ and $\gamma_\isrc{}$ with informative priors used for
these parameters. We note that our framework can also handle the case
between 1) and 2), including some balance between global and
individual source parameters, and this is discussed further in
Section~\ref{sec:discussion}.

\begin{figure*}
  \centering
  \includegraphics[width=\textwidth]{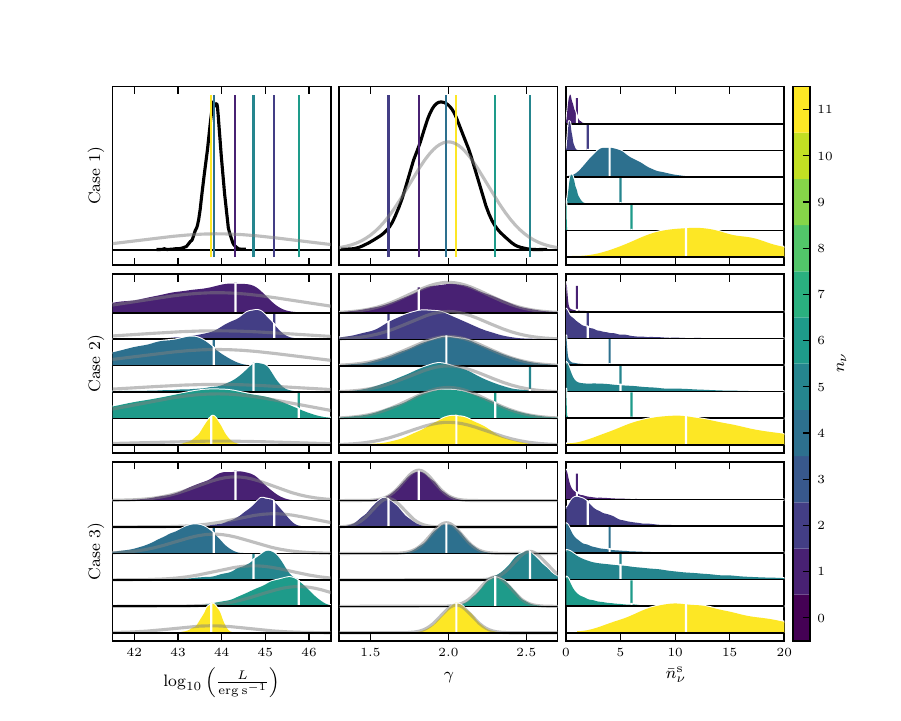}
  \caption{The marginal posteriors for $L$, $\gammaps{}$ and
    $\Nnupsiexp{\isrc}$ for all 6 point sources, shown by the
    different colours. The upper, middle and lower rows show the
    results for cases 1), 2) and 3), respectively. The true values are
    indicated as solid vertical lines and the priors are shown in
    grey, where relevant.}
  \label{fig:L_n_gamma_multi_source}
\end{figure*}

For the joint fits, all information shown in
Section~\ref{sec:single_source} is available and can be analysed
independently for all sources. The difference between fitting all
sources separately is that the joint fit allows the results of one
source to influence the others, according to the details of the model
for their population. For brevity, we focus here on few key results
that highlight the implications of the three example cases introduced
above. Fig.~\ref{fig:L_n_gamma_multi_source} shows the marginal
posterior distributions for the point source luminosity, $L$ and the
expected number of events from each source, \Nnupsexp{}, for each of
the three studied cases.

For the first case of shared $L$ and \gammaps{}, we see that the
source with most events, source \#1, drives the values of these
parameters.  As this source has a relatively low $L$\footnote{Despite
  having a low $L$, as source \#1 is the closest source, it still has
  the highest number of detected events in the simulated data set.},
\Nnupsiexp{\isrc} is reduced for all sources other than source \#4,
which has a similar $L$. For this case, \gammaps{} is relatively
unconstrained due to the low event numbers, so there is only a slight
information gain of the posterior relative to the prior. The overall
impact is the underestimation of \Nnupsiexp{\isrc} for harder sources,
such as sources \#3, \#5 and \#6.

For the case of individual free source parameters, $L_\isrc$ and
$\gamma_\isrc$, the variety of the sources is better captured. While
the $L_\isrc{}$ and $\Nnupsi{\isrc}$ posteriors are all consistent
with the corresponding true values, the distributions remain
relatively unconstrained for sources \#2, \#4 and \#6. The
$\gamma_\isrc$ posteriors show some shift away from the prior for
sources \#5 and \#1, but are otherwise rather similar to the posterior
from the shared parameter case above.

Finally, we investigate using informative priors for the individual
$L_\isrc{}$ and $\gamma_\isrc$. Cases 1) and 2) both use shared priors
for the source parameters, as described in Table~\ref{tab:priors}. We
now suppose that further information is available on the $L_\isrc$ and
$\gamma_\isrc$ for the individual sources. This could be in the form
of, e.g.,~neutrino observations of other sources, multi-messenger
information, or theoretical predictions. We use log-normal priors for
$L_\isrc$, centred on the true values and with a width of 2.0. For the
$\gamma_\isrc$, we use normal priors centred on the true values with
$\sigma_\gamma = 0.1$. While informative, these priors still allow for
the possibility that $\Nnupsiexp{\isrc{}} = 0$ for all sources, and
the allowed parameter ranges leave room for the data to drive the
posterior results. We see that in this case, we have better
reconstruction of both $L_\isrc$ and \Nnupsiexp{\isrc{}} for all
sources, but the most impact is seen for sources \#2, \#4, and \#6,
which were the least constrained in case 2). The $\gamma_\isrc$
posteriors mostly follow the priors, with no strong additional
constraints from the data.

\begin{figure}
  \centering
  \includegraphics[width=\columnwidth]{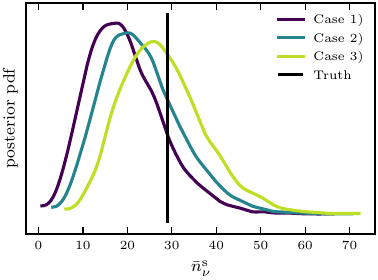}
  \caption{The marginal posterior distributions for \Nnupsexp{} from
    all 6 point sources for the three example cases discussed in
    Section~\ref{sec:multi_source}.}
  \label{fig:Nex_comp}
\end{figure}

We further summarise the above results for the three example cases by
comparing the results for the total number of expected point source
events across all sources, \Nnupsexp{}, as shown in
Fig.~\ref{fig:Nex_comp}. We see that assuming the $L$ and \gammaps{}
are shared by all sources as in case 1), we underestimate the total
contribution of these point sources as $\sim 1/2$ of the true
value. For case 2) we improve the result, but still underestimate the
contribution as only $\sim 3/4$ of the truth and the extra information
in case 3) is necessary to recover the contribution from all point
sources. As $\Pr(\Nnupsexp{} \geq 1 | \Edet{}, \dirdet{}) = 1.0$ for
all cases, it makes more sense to state the \Nnupsexp{} threshold for
which this expression is $> 0.95$. For cases 1), 2) and 3) this is 7,
9 and 14, respectively. We can interpret this is a confident detection
of this many events from all point sources included in the fit.

We give the $\Pr(\Nnupsiexp{\isrc{}} \geq 1 | \Edet{}, \dirdet{})$
values for each source considered in
Table~\ref{tab:p_det_multi_src}. Interestingly, for case 1), both
sources \#1 and \#4 are detectable according to the definition in
\eqref{eq:discovery_nex} with $\alpha=0.95$, but only source \#1
remains detectable for cases 2) and 3). This can be understood as
sources \#1 and \#4 have very similar $L_\isrc$ and $\gamma_\isrc$, so
it is beneficial to assume shared parameters in this case, especially
given the strength of source \#1. However, the overall detectability
of all sources is better served by considering their individual
properties, and the value of
$\Pr(\Nnupsiexp{\isrc} \geq 1 | \Edet{}, \dirdet{})$ increases
significantly for all sources other than \#4.

\begin{table}[ht]
  \centering
  \begin{tabular}{cccc}
    \toprule
    \multirow{2}{*}{Source} & \multicolumn{3}{c}{$\Pr(\Nnupsiexp{\isrc{}} \geq 1 | \Edet{}, \dirdet{})$} \\
    \cmidrule{2-4}
                            & Case 1 & Case 2 & Case 3 \\
    \cmidrule{1-4}
    \#1 & 0.998 & 0.998 & 1.000 \\
    \#2 & 0.000 & 0.098 & 0.566 \\
    \#3 & 0.158 & 0.777 & 0.859 \\
    \#4 & 0.989 & 0.305 & 0.568 \\
    \#5 & 0.007 & 0.661 & 0.769 \\
    \#6 & 0.110 & 0.316 & 0.436 \\
    \bottomrule
  \end{tabular}
  \caption{Impact of the three different modelling assumptions
    described in the text on the detectability of all sources
    considered.}
  \label{tab:p_det_multi_src}
\end{table}

The event--source association probabilities generally reflect the
results discussed above. It is worth mentioning that source \#5
produces an event with $\Edet{} = 1.3$~\si{\PeV} in our simulation,
which has an association probability of 0.92, 0.95 and 0.99 for cases
1), 2) and 3), respectively. As this is a faint source with only 2
simulated events and more extreme properties, it makes sense that the
association probability increases as we allow the sources to have
independent parameters and include more prior information.

\subsection{Comparison with \texttt{SkyLLH}}
\label{sec:comp_skyllh}

\begin{figure*}
  \centering
  \includegraphics[width=0.45\textwidth]{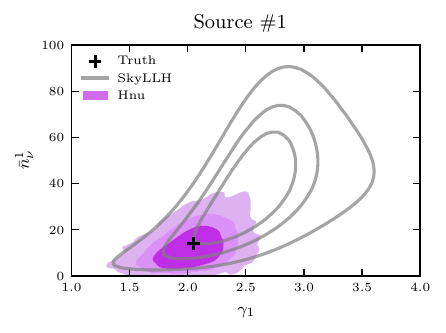}
  \includegraphics[width=0.45\textwidth]{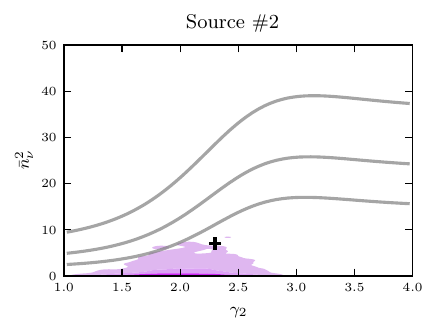}
  \includegraphics[width=0.45\textwidth]{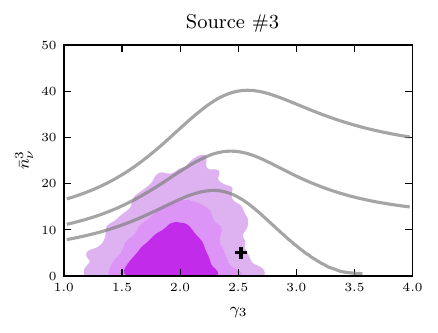}
  \includegraphics[width=0.45\textwidth]{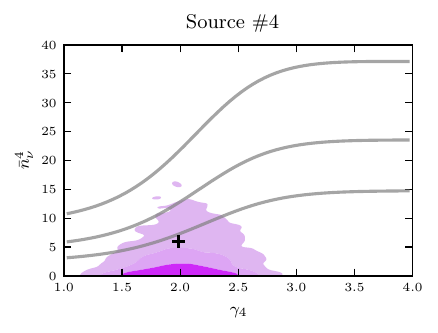}
  \includegraphics[width=0.45\textwidth]{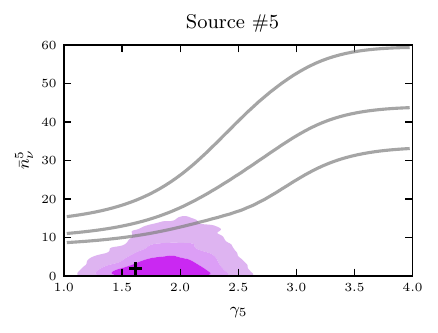}
  \includegraphics[width=0.45\textwidth]{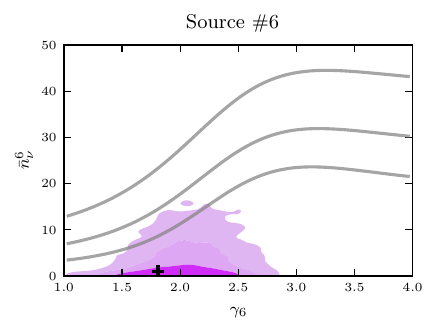}
  \caption{Comparison of the analysis of the simulated sources with
    \texttt{SkyLLH} and the \texttt{hierarchical\_nu} framework
    presented in this work. The grey contours show the 68, 90, and 99\%
    confidence intervals computed by assuming Wilk's theorem with 2
    degrees of freedom. The purple shaded contours show the Bayesian
    68, 90, and 99\% credible regions of highest posterior density.}
  \label{fig:comp_skyllh}
\end{figure*}

We use \texttt{SkyLLH} \citep{Wolf:2019Pu,Bellenghi:20230u} to analyse
our the sources in our simulated data set individually, showing the
results for the source properties in Fig.~\ref{fig:comp_skyllh}. We
compare these results to our results using our
\texttt{hierarchical\_nu} framework for case 2) described in
Section~\ref{sec:multi_source}, which is the most similar to
considering all sources individually. We note that the contours shown
in Fig.~\ref{fig:comp_skyllh} represent frequentist confidence
intervals and Bayesian credible regions for the \texttt{SkyLLH} and
\texttt{hierarchical\_nu}, respectively, which have different
underlying definitions. For all sources the results are consistent
with the true values and we see that our framework leads to stronger
constraints on the source properties. For sources \#2 and \#3, where
the truth is on the edge of the posterior distributions from
\texttt{hierarchical\_nu}, the combination of the weakly informative
prior centred on $\gamma = 2$ with the softer true $\gamma$ and the
energy distribution of events found in this particular realisation of
the simulated data set impacts the results, as expected. In
Section~\ref{sec:multi_source}, we show how different connections
between sources and more informative priors can be used to mitigate
this impact in a model-dependent way.

\section{Discussion}
\label{sec:discussion}

The results shown in Section~\ref{sec:single_source} and
\ref{sec:multi_source} demonstrate the validity of our method and
introduce the possible analyses that can be performed within this
framework. In addition to parameter estimation, we address how
probabilities of interest can be quantified, such as the event--source
association probability and the probability that a source or
population contributes at least \Nnupsexp{} events to the data. The
most relevant output will typically depend on the details of the
application to different physical scenarios. For example, the
event--source association probabilities are likely most useful for
studying high-energy neutrino alert events or analysing the impact of
different source spectral models on possible associations. On the
other hand, the overall expected event contribution is a direct way to
investigate the detectability of sources and their populations that
naturally includes the relevant uncertainties. Example applications
could include characterising a population of hard-spectrum sources
that TXS~0506+056-like sources
to~\citep{Buson:2022ps,Buson:2023pf,Bellenghi:2023kd}, or constraining
the contribution from a population of NGC~1068-like
sources~\citep{Glauch:2023Qy,Saurenhaus:2023De}. With our framework,
we aim to provide a consistent setting for application to these
different cases.

In Section~\ref{sec:multi_source}, we also illustrate the power that
including more information into the analysis can have when trying to
detect a number of weak sources in the data. Even for events with a
relatively poor energy resolution, we see in Fig.~\ref{fig:Nex_comp}
that including prior information on the spectral index makes a
significant difference when estimating the contribution of these point
sources. For a non-detection, this equates to stronger constraints on
the proposed model, allowing us to make the most of the available data
in either case.
 
Several recent works in the field of multi-messenger astrophysics
explore Bayesian approaches for individual source--event associations
(see e.g. \citealt{Ashton:2018gq}, \citealt{Bartos:2019mf} and
\citealt{Veske:2021kd}, \citealt{Kowalski:2021pq}). These methods
still frame the problem as a hypothesis test, preferring Bayes factors
and odds ratios to $p$-values, or using the Bayes factor as a test
statistic in order to draw conclusions. Here, we focus more on
addressing the questions of interest via parameter estimation rather
than model comparison. The foreseen workflow is to be able to develop
and refine models that are consistent with the data, while exploring
their implications. In this workflow, model rejection is a subset of
the possible outcomes.

Frequentist hypothesis testing methods are typically used in searches
for neutrino point sources. In particular, likelihood ratio methods as
introduced in \citet{Braun:2008bg} and implemented in \texttt{SkyLLH}
have been used in IceCube analyses to find evidence for TXS~0506+056,
NGC~1068 and the Galactic plane as neutrino sources
\citep{IceCube:2018dnn,IceCube:2018kh,Abbasi:2022sw,Abbasi:2023pd}. We
see our Bayesian approach as complemetary to the standard methods in
that the focus is on the evaluation and charaterisation of source
models rather than the rejection of the background-only
hypothesis. Due to the different definition of probability in
frequentist and Bayesian statistics and the different goals of these
two methods, it is non-trivial to directly compare their
performance. As such, we tend to make more qualitative comparisons in
this work and highlight the complementary features below.

With our definition of detection and the realisation of the simulated
data set studied, we can detect source \#1 with only weakly
informative priors, as well as source \#4 when assuming shared source
parameters as in case 1) of Section~\ref{sec:multi_source}. Source \#5
has one very high energy event and is therefore detectable for case
3). Sources \#2, \#3 and \#6 cannot be detected independently as they
are softer (\#2 and \#3) or happen to only produce lower energy events
in this simulated data set (\#6), but their contribution to the
population can be recovered when including additional information, as
shown in Fig.~\ref{fig:Nex_comp}.

We recognise that our joint source fits are conceptually similar to
the ``stacking'' technique that has been used in previous point source
searches to increase sensitivity to particular source modelling
assumptions (e.g. \citealt{Glauch:2023Qy}). Case 1) is most similar to
a distance-weighted stacking analysis and case 3) is similar to a
flux-weighted stacking analysis. However, an important difference is
that the priors and modelling that we use are set up in a way such
that relevant uncertainties are included and the data can overrule the
prior in more informative cases. This modelling, together with the
interpretation offered by the Bayesian approach results in a more
flexible analysis.

Cases 1) and 2) explore the possibility that all sources are the same
(``complete pooling'') and that all sources and independent (``no
pooling''), respectively. Neither of these assumptions are completely
realistic, and in practice we expect some balance between global and
individual source properties (``partial pooling''), as modelled in our
simulation. We see for case 2) that the neutrino data alone does not
contain enough information to significantly constrain the population
hyperparameters (the shape of the luminosity function and spectral
index distribution). However, the future increase in performance
expected from planned experiments mentioned in
Section~\ref{sec:introduction} will provide the data sets necessary
for the ``partial pooling'' case to be leveraged, and we plan to
explore the impact of these new possibilities in upcoming work.

Thanks to the implementation of our statistical model with \stan{}'s
Hamiltonian Monte Carlo algorithm, we are able to perform fits with
over $\sim 7000$ free parameters in Section~\ref{sec:multi_source},
where the majority of the free parameters are the latent \E{} of the
events. In principle, these latent \E{} parameters could be
marginalised over in the likelihood to speed up the fits, but as
discussed in Section~\ref{sec:interpretation} and demonstrated in
Section~\ref{sec:single_source}, these parameters add to the
interpretability of the results. By including these free parameters
here, we also demonstrate that it is relatively straightforward to add
further model complexity in terms of parameters for both the source
modelling and detector modelling. Large numbers of free parameters can
be challenging for optimisers currently implemented in
\texttt{SkyLLH}, as it was not designed to fit the latent \E{}
parameters or more complex source models. Markov chain Monte Carlo
methods, such as those used here, could also be implemented in a
frequentist setting to address these challenges.

Another way in which we include more complexity in this work is that
we model the atmospheric and diffuse background components, while
keeping the total number of expected events conserved. In this way, if
events are fit to point sources, the diffuse astrophysical and/or
atmospheric components are reduced to compensate. This approach is
interesting when considering the bigger picture of possible neutrino
sources, with competing source populations expected (e.g.,
\citealt{Bartos:2021sj}), in addition to contributions from the
Galactic plane.

A challenge of the frequentist hypothesis testing approach arises when
testing multiple hypotheses, necessitating a trial correction
factor. Consequently, source lists for studies have to be limited or
the discovery threshold raised, leading to less use of the data and
decreased sensitivity. Furthermore, it is non-trivial to keep track of
trial factors across independent analyses of the same data. While the
Bayesian methods used here do not guarantee a certain false positive
rate, the structure included through the priors and hierarchy of our
model naturally tends to mitigate the effects of outliers
\citep{Gelman:2009hw}. Should a certain false positive rate be
desired, it is possible to calibrate the probability thresholds
introduced in Section~\ref{sec:interpretation} through repeated
simulation and fits, as detailed in \citet{betancourt2018calibrating}.

Along with guaranteed coverage and false positive rates, one
complementary aspect of the standard approach is that by focusing on
rejecting background, unexpected signals can be identified even if the
alternate hypothesis is not well-specified. In our approach,
mismodelling can be also be studied with PPCs and used to improve
models for a better match with the data. Additionally, computational
challenges mean that our approach is currently better suited to
building and testing specific point source models rather than
performing an uninformed scan across the whole sky. Our implementation
in \texttt{Stan} includes within-chain parallelisation of the
likelihood evaluation that can be scaled according to the available
computational resources. As a benchmark, for point sources near the
Equator where event rates are the largest, fits using 10 years of data
selected in an ROI of \ang{5} radius may take up to 2 hours on 144
threads.

\section{Conclusions}
\label{sec:conclusions}

We present a hierarchical Bayesian approach to searching for point
sources of astrophysical neutrinos. Our method is an alternative to
existing frequentist approaches, with a focus on the characterisation
of sources and interpretability of results.

We demonstrate our approach through application to a simulated data
set containing 6 weak point sources hidden in typical
backgrounds. Even for low event numbers of
${\Nnupsi{\isrc{}} \leq 14}$, we are able to recover the constribution
of the strongest source and provide constraints on its luminosity and
spectral index. The contribution of the remaining sources can also be
inferred by leveraging similar source properties and more informative
priors for their spectral shape. These results show the potential gain
of more complex and flexible modelling when studying weak sources that
is relevant to the expected signals from current neutrino source
candidates, such as blazars or Seyfert galaxies.

We plan to apply our framework to the existing public data sets,
investigate the impact of more complex source modelling, and study the
information gain from expected future data sets. Our framework is
written in a modular way, such that it can also be extended to other
event types and detectors, providing a useful open-source tool to the
community for the evaluation of different astrophysical models and
detector configurations.

\begin{acknowledgments}

  The authors express their thanks to Chiara Bellenghi and Martin Wolf
  for very helpful discussions about the used data set, as well as to
  Lena Saurenhaus for contributing to the production of
  Fig.~\ref{fig:sim_summary}. The authors are also grateful to Allen
  Caldwell and Daniel Mortlock for feedback on an early version of
  this manuscript.
  
  Julian Kuhlmann acknowledges support from the DFG through the
  Sonderforschungsbereich SFB 1258 ``Neutrinos and Dark Matter in
  Astro- and Particle Physics" (NDM). Francesca Capel acknowledges the
  financial support from the Excellence Cluster ORIGINS, which is
  funded by the Deutsche Forschungsgemeinschaft (DFG, German Research
  Foundation) under Germany's Excellence Strategy - EXC-2094-390783311
  at the early stage of this work.

\end{acknowledgments}

\vspace{5mm} \facilities{Computations were performed on the HPC system
  Raven at the Max Planck Computing and Data Facility.}

\software{\texttt{stan} and \texttt{cmdstanpy} \citep{Stan:2023pf},
  \texttt{astropy} \citep{2013A&A...558A..33A,2018AJ....156..123A},
  \texttt{numpy} \citep{numpy}, \texttt{matplotlib}
  \citep{Hunter:2007}, \texttt{scipy} \citep{2020SciPy-NMeth},
  \texttt{h5py} \citep{collette_python_hdf5_2014,h5py_7560547},
  \texttt{arviz} \citep{Kumar_ArviZ_a_unified}, \texttt{seaborn}
  \citep{Waskom2021}.}

\appendix

\section{Detector model}
\label{sec:neutrino_detection}

Using the information provided in the publicly available IRFs
\citep{icecube_data}, we can calculate the number of expected events
in a sample as
\begin{equation}\label{eq:app_nex}
  \Nnutotexp{} = T \int_{\Emin{}}^{\Emax{}} \dd{\E{}}\int_{\Edetmin{}}^{\Edetmax{}}\dd{\Edet{}}
  \int_{S_2} \dd{\dir{}} \int_\mathrm{ROI} \dd{\dirdet{}}
  \frac{\dd[4]{\Nnu{}}}{\dd{\E{}}\dd{t}\dd{A}\dd{\dir{}}}
  \Aeff{}(\E{}, \dir{})\,
  \IRF{\Edet{}\vert\E{}, \dir{}}\,\IRF{\dirdet{}\vert\dir{}, \E{}, \Edet{}},
\end{equation}
where $T$ is the total observation time, ROI is the region of interest
and $S_2$ the surface of a sphere. The integrals over solid angle
reflect the modelling of the sources and the cuts on reconstructed
direction, respectively. In fits, the angular resolution is modelled
using a Rayleigh distribution embedded on a sphere, for a derivation
see \citet{dissertation_glauch}. In simulations, $\angerr{}$ is
sampled from the provided IRFs.

Due to the required differentiability of the model likelihood with
respect to the neutrino energy, we have to interpolate the provided
energy resolution, which is provided as histograms in reconstructed
energies, covering half a decade of neutrino energy, split over three
declination bands.  For each histogram $\Pr(\Edet{}\vert\E{})$ we fill
zero-entries in between non-zero entries by interpolation and extend
the empty flanks by a steep power-law with index $\pm 15$. After
renormalising to unity, we stack the histograms' logarithm along a new
axis of logarithmic neutrino energy and create a two-dimensional
spline representation.  Evaluations of this spline at each \Edet{}
over a dense grid of $\log_{10}(\E{})$ is handed over to \stan{} for
interpolation. Similarly, we interpolate the logarithm of effective
area over logarithmic energies.

As the provided instrument response is only valid for events
originating as neutrinos we are restricted to analysis in the Northern
hemisphere, as in the Southern hemisphere the event rate is dominated
by atmospheric muons.

\bibliography{hierarchical_nu}{} \bibliographystyle{aasjournal}

\end{document}